\documentclass[fleqn,10pt]{wlscirep} 
\usepackage{amsmath,graphicx}
\usepackage{tikz}
\usepackage[all,poly]{xy}
\usepackage{multirow}

\usepackage{amsmath}
\usepackage{amsfonts}
\usepackage{textcomp}
\usepackage{gensymb}
\usepackage{lineno,hyperref}
\usepackage{subcaption}
\usepackage{xcolor}
\usepackage{chngcntr}
\usepackage{natbib}
\usepackage{enumitem}
\usepackage{multirow}
\counterwithout{figure}{subsection}
\usepackage[final]{changes}
\newcommand{\stkout}[1]{\ifmmode\text{\sout{\ensuremath{#1}}}\else\sout{#1}\fi}
\setdeletedmarkup{\stkout{#1}}
\usepackage{mathtools}
\usepackage[linesnumbered, ruled, vlined]{algorithm2e}

\title{Quantitative Imaging and Automated Fuel Pin Identification for Passive Gamma Emission Tomography}

\author[1,*]{Ming Fang}
\author[2]{Yoann Altmann}
\author[3]{Daniele Della Latta}
\author[4]{Massimiliano Salvatori}
\author[1]{Angela Di Fulvio}

\affil[1]{Department of Nuclear, Plasma, and Radiological Engineering, University of Illinois at Urbana-Champaign, Urbana, IL 61801, United States}
\affil[2]{School of Engineering and Physical Sciences, Heriot-Watt University, Riccarton, Edinburgh, EH14 4AS, United Kingdom}
\affil[3]{Fondazione Toscana Gabriele Monasterio, Via Giuseppe Moruzzi, 1, 56124 Pisa PI, Italy{ (now with Terarecon Inc., 4309 Emperor Blvd, Suite 310 Durham, NC 27703, United States)}}
\affil[4]{Fondazione Toscana Gabriele Monasterio, Via Giuseppe Moruzzi, 1, 56124 Pisa PI, Italy{ (now with Kymamed Srls, Viale Roma 208 54100 Massa Italy)}}

\affil[*]{mingf2@illinois.edu}

\begin{abstract}
Compliance of member States to the Treaty on the Non-Proliferation of Nuclear Weapons is monitored through nuclear safeguards. {The Passive Gamma Emission Tomography (PGET) system is a novel instrument developed within the framework of the International Atomic Energy Agency (IAEA) project JNT 1510, which included the European Commission, Finland, Hungary and Sweden. The PGET is used for the verification of spent nuclear fuel stored in water pools.} Advanced image reconstruction techniques are crucial for obtaining high-quality cross-sectional images of the spent-fuel bundle to allow inspectors of the IAEA to monitor nuclear material and promptly identify its diversion. In this work, we have developed a software suite to accurately reconstruct the spent-fuel cross sectional image, automatically identify present fuel rods, and estimate their activity. Unique image reconstruction challenges are posed by the measurement of spent fuel, due to its high activity and the self-attenuation.
 While the former is mitigated by detector physical collimation, we implemented a linear forward model to model the detector responses to the fuel rods inside the PGET, to account for the latter. The image reconstruction is performed by solving a regularized linear inverse problem using the fast-iterative shrinkage-thresholding algorithm (FISTA). We have also implemented the traditional filtered back projection (FBP) method based on the inverse Radon transform for comparison and applied both methods to reconstruct images of simulated mockup fuel assemblies. Higher image resolution and fewer reconstruction artifacts were obtained with the inverse-problem approach, with the mean-square-error (MSE) reduced by 50\%, and the structural-similarity (SSIM) improved by 200\%. We then used a convolutional neural network (CNN) to automatically identify the bundle type and extract the pin locations from the images; the estimated activity levels finally being compared with the ground truth. The proposed computational methods accurately estimated the activity levels of the present pins, with an associated uncertainty of approximately 5\%.  
\end{abstract}

\begin{document}

\flushbottom
\maketitle
%
%
\thispagestyle{empty}

\section*{Introduction}
Under the Treaty on the Non-Proliferation of Nuclear Weapons (NPT)~\cite{NPT1970} and other treaties against nuclear proliferation, the International Atomic Energy Agency (IAEA) is entrusted to verify all nuclear materials under the control of the State with safeguards agreements in force with the Agency.
International safeguards include all the technical measures put in place by the IAEA to verify that each State Parties complies with the aforementioned agreements. 

Most safeguards methods are based on the assay of nuclear materials through measurements of ionizing radiation emitted by the sample. Safeguards systems are undergoing an important modernization effort~\cite{safechal}.
One example of such effort is the development of the PGET systems for the inspection of spent fuel in water pools{~\cite{honkamaa2014prototype,white2018application,PGET}}. Traditionally, an inspection of spent fuel in pools was performed using the FORK detector, which encompasses, in its standard version, one ionization chamber to measure gamma rays and two neutron-sensitive fission chambers. The FORK detector, however, exhibits low sensitivity to the diversion of fuel pins, being able to detect the diversion of 50\% or more of the fuel pins in the assembly under inspection~\cite{forkDet}.
A PGET was recently developed to increase the sensitivity to a variety of fuel diversion scenarios~\cite{PGET}. The software to analyze PGET images can be improved~\cite{iaea2019challenge} to reduce user intervention and overall perform faster and more reliable monitoring of spent fuel and therefore implement the ultimate safeguards objective, \textit{i.e.}, the prompt detection of nuclear material diverted from peaceful uses.

The PGET of spent fuel poses some unique challenges, compared to industrial or medical tomographic imaging, which are due to the high activity of the sources (a single-pin activity is of the order of $10^{13}$~Bq) and the high self-attenuation of the fuel pins. While the former can be mitigated using collimated detectors, the latter needs to be addressed by using specific imaging algorithms.{ Previous work has shown that the reconstruction quality could be significantly improved by applying the attenuation correction{~\cite{1930-8337_2020_2_317}}. In this work, we further corrected for the gamma ray down-scattering in the energy window of interest.}
We have implemented and integrated computational methods based on proximal gradient, physics-informed Monte Carlo sampling, and machine learning to (1) improve the PGET image quality{,} (2) automatically identify missing pins{, and (3) quantify the pin activities}. As we will show, the automated identification of missing pins relies heavily on the image quality, and therefore the three tasks are inherently intertwined.



\section*{Results}

\subsection*{Simulated Sinograms}
\begin{figure}[!htbp]
\captionsetup{font=footnotesize}
    \centering
    \includegraphics[width=\linewidth]{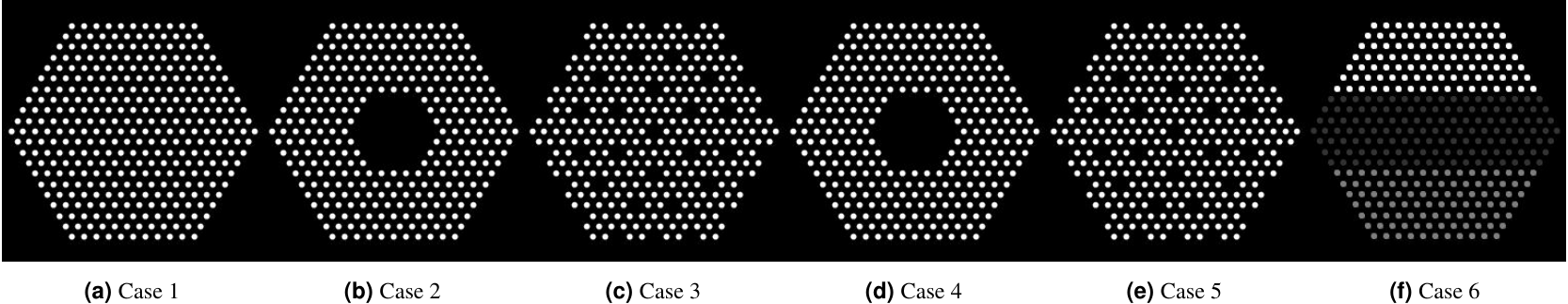}
    \caption{Actual pin distributions in the simulated fuel assemblies (ground truth).{ Pin activity is proportional to the brightness. In Case 1, the mock-up fuel assembly hosted 331 ${}^{60}$Co pins of 8.879~g/cm${}^3$ density, which emitted 1.17 and 1.33~MeV gamma rays. In Cases 2 and 3, 10\% ${}^{60}$Co pins pins were removed at the center or uniformly. In Cases 4 and 5, 10\% ${}^{60}$Co pins pins were replaced by depleted uranium pins of the same size but higher density (10.4~g/cm${}^3$). In Case 6, we decreased the activities of the middle pins by 80\% and bottom pins by 50\%.} }
    \label{fig: pin_distri}
\end{figure}

We simulated the measurement of six fuel pin configurations in a mock-up water-water energetic reactor (VVER) fuel assembly using MCNP 6.2~\cite{goorley2013initial}, as shown in Fig.~\ref{fig: pin_distri}. In Cases 2 and 3, we mimicked the scenario where fuel pins were missing. In Cases 4 and 5, {we used depleted uranium to replace cobalt because it resulted in the highest attenuation among different potential replacement materials~\cite{di2019neutron}, which would allow us to examine the algorithm's capability of \replaced{distinguishing replaced}{identifying missing} pins\deleted{ in the worst-case scenario}. }In Case 6, we {wanted to explore whether the algorithm is able to render an accurate image of a space-dependent activity distribution.} Details of the simulated PGET model are included in {``Methods-Simulation Methods"} section.

\begin{figure}[!htbp]
\captionsetup{font=footnotesize}
\centering
  \includegraphics[width=.5\textwidth]{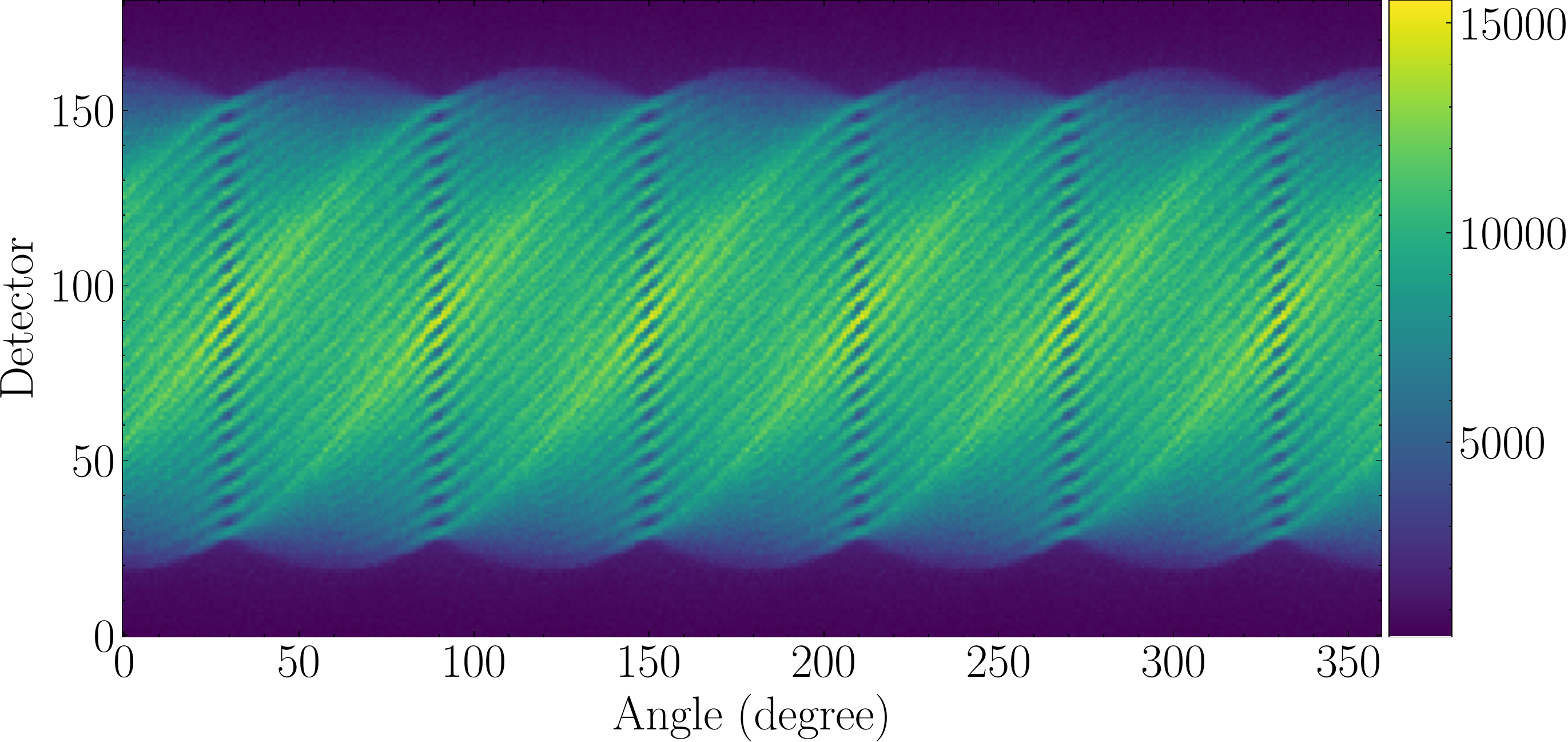}
  \caption{Simulated sinogram of the fully-loaded bundle. The color scale represents the detected photon counts.{The periodicity of the singogram resulted from the hexagonal symmetry of the simulated fuel assembly.}}
  \label{fig:sinos}
\end{figure}

{The detection system we simulated is the so-called PGET prototype device as described by~\cite{white2018application}. }The detection unit consisted of \replaced{two collimated CdTe detector arrays on opposite sides of the fuel assembly}{two collimated CdTe detector arrays on the top and bottom sides of the fuel assembly}, each encompassing 91 detectors. The detector arrays rotated and scanned the fuel bundle in steps of 1$\degree$, which generated a $182\times360$ sinogram, as shown in Fig.~\ref{fig:sinos}. We simulated {$5\times 10^9$} {NPS} (number of {particle histories}) in each case and detected the photons in the 700-1500~keV energy window. 
Approximately {15,000} maximum counts in one pixel were achieved, comparable to the counts in an actual PGET measurement.

\subsection*{Inverse Approach}
\begin{figure}[!htb]
 \captionsetup{font=footnotesize}
    \centering
    \includegraphics[width=0.4\linewidth]{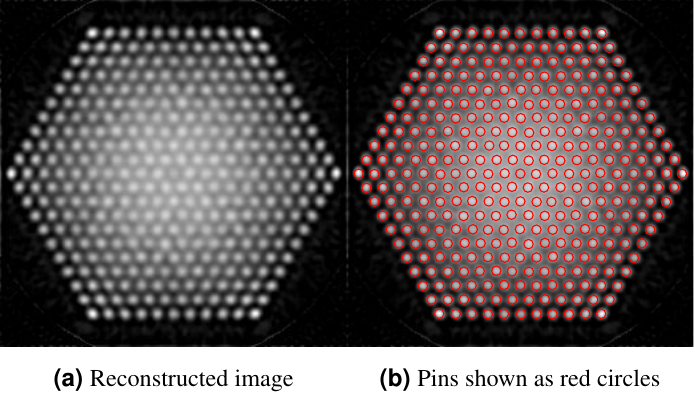}
    \caption{Image reconstruction and pin identification using the simple response matrix in Case 1.}
    \label{fig: init}
\end{figure}

\begin{figure}[!htb]
  \captionsetup{font=footnotesize}
    \centering
    \includegraphics[width=.23\linewidth]{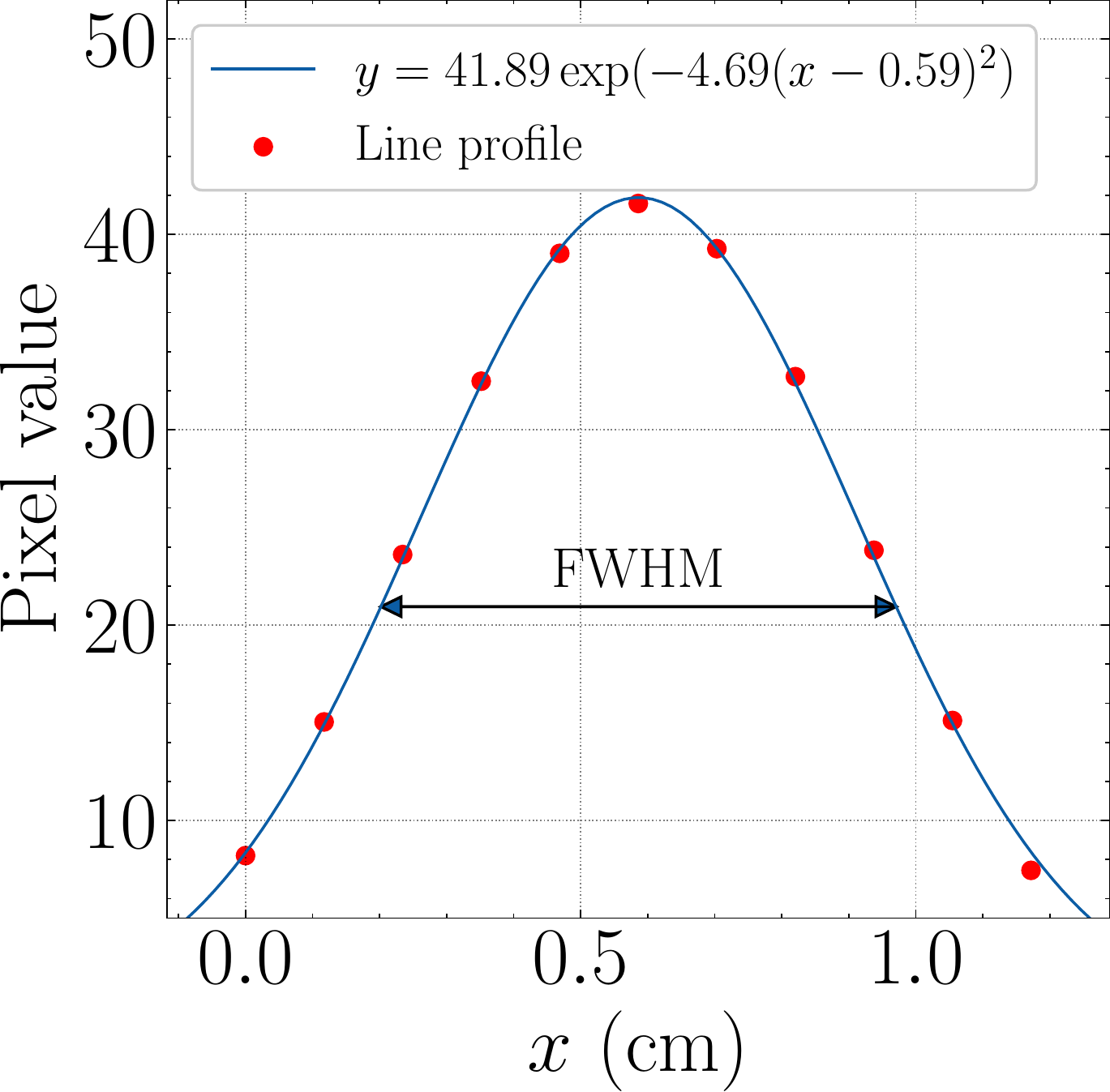}
    \caption{Gaussian fit of the average line profile.{ FWHM of the fitted Gaussian is shown as the arrow. The 11 pixels around the pin center shown as the red points were used in the fitting. The coefficient of determination $R^2$ of the fit was 0.9991.}}
    \label{fig:init_gaus_fit}
\end{figure}

We have developed a systematic approach that allows us to reconstruct high-quality images from individual sinograms, identify the pin locations, and quantify the pin activity levels.{ The image of fuel pins was reconstructed by solving a linear inverse problem. The region of interest was divided into $182 \times 182$ pixels. We calculated the detector array response to a source of unit activity inside each pixel, forming the system response matrix. For an unknown fuel assembly, the measured sinogram is a linear combination of the calculated responses, with the coefficients being the source strength of each pixel. In this way, the image reconstruction problem was converted into a linear inverse problem, with both the simulated sinogram and calculated response matrix as inputs.

First, we performed a simple image reconstruction to obtain an initial estimate of pin locations and pin radius. We calculated the system response matrix using a deterministic ray-tracing method, assuming that the scattering of gamma rays can be neglected~\cite{fanginmm2020}. The reconstructed image of Case~1 is shown in Fig.~\ref{fig: init}a. Inner pixels are brighter than the outer ones on the image, indicating an overestimation of the activity of the pins. This is because the contribution of scattered photons to the system response cannot be neglected for inner pixels, due to the heavy attenuation of unscattered photons. Nevertheless, using this image, we can determine the center of all possible pins as the centroids of the bright regions on the reconstructed image. In this step, we would allow some slackness in pin identification, since a more accurate reconstruction will be performed later. The identified pins are shown in Fig.~\ref{fig: init}b. Horizontal and vertical line profiles passing through the center of each pin were extracted from the image, and we fitted a Gaussian to the average of all profiles, shown in Fig~\ref{fig:init_gaus_fit}. The FWHM (full width at half maximum) of the fitted Gaussian was 0.769~cm, which was taken to be the pin {diameter}.}

\begin{figure}[!htbp]
\captionsetup{font=footnotesize}
    \centering
    \includegraphics[width=\linewidth]{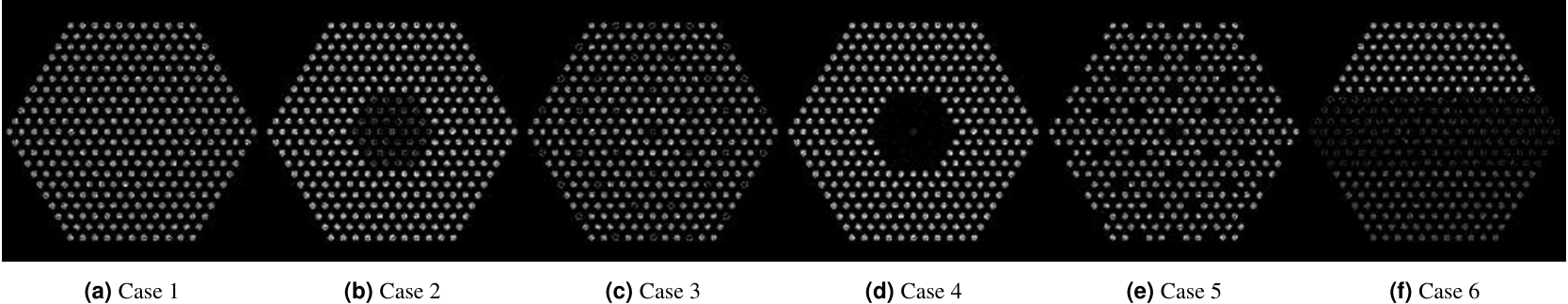}
    \caption{Image reconstructed using the inverse approach.{ Good contrast between the pin-present region and pin-absent region was achieved. No severe artifacts were observed on the reconstructed images.}}
    \label{fig: inv_recon}
\end{figure}
\begin{figure}[!htbp]
\captionsetup{font=footnotesize}
    \centering
    \includegraphics[width=\linewidth]{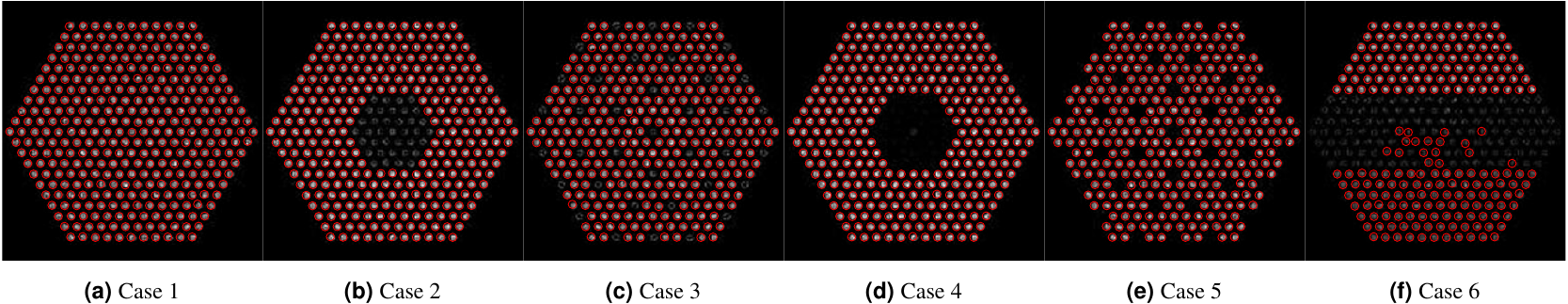}
    \caption{Pin identifications based on the inverse approach. Identified pins are shown as red circles.{ For Cases 1 to 5, we achieved 100\% classification accuracy; for Case 6, we achieved 100\% classification accuracy for the pins of medium and high activity level.}}
    \label{fig: inv_pid}
\end{figure}

{With the pin locations and pin radius known, we could create a material map of the fuel assembly. We then \replaced{implemented an accelerated Monte-Carlo algorithm}{performed a Monte Carlo simulation} to calculate the system response matrix, accounting for both the absorption and scattering interaction of photons in the fuel assembly. The images reconstructed using the new response matrix are shown in Fig.~\ref{fig: inv_recon}. The images were then fed to a convolutional neural network to perform pin identification. The results are shown in Fig.~\ref{fig: inv_pid}. For Cases 1 to 5, we achieved 100\% classification accuracy, due to the high quality of reconstructed images. We created a histogram of pin activities, as shown in Fig.~\ref{fig: inv_act}. In Cases 1-5,  {the standard deviation  of activity estimation ranged from 3\% to 6\%, and }the mean of the pin activity distribution deviated from the ground truth by less than 4\%. In Case 6, we successfully identified all pins, except those with the lowest activity level (Group 1) shown in \added{Fig.~\ref{fig: inv_pid}f and }{Fig.~\ref{fig: inv_act}f}. The activity of pins in Group~2 and Group~3 was accurately reproduced with a relative error below 1\%. The inverse approach is detailed in the ``Methods-Image Reconstruction Methods" section.}

\section*{Discussion}
\begin{figure}[!htbp]
\captionsetup{font=footnotesize}
    \centering
    \includegraphics[width=.5\linewidth]{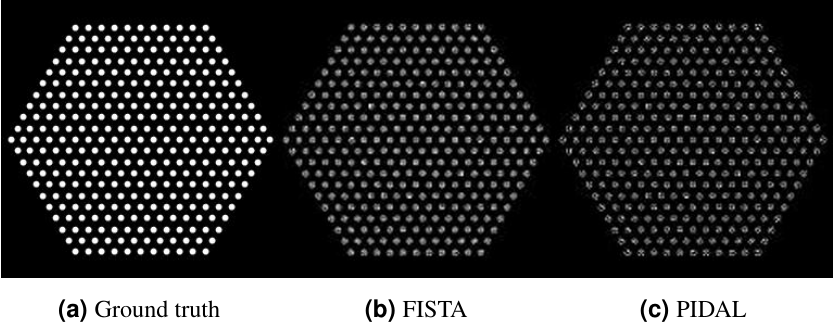}
    \caption{Comparison of FISTA and PIDAL reconstruction{ to the ground truth. The two methods led to visually similar results, though the PIDAL reconstruction resulted in a sparser image, compared to FISTA}.}
    \label{fig:fistavspista}
\end{figure}

The image reconstruction problem can be formulated into different linear inverse problems based on the selected model of observation noise. As detailed in ``Inverse Problem Approach", we formulated two inverse problems with the Gaussian noise model and Poisson model, which can be solved using FISTA { (fast-iterative shrinkage-thresholding algorithm)~\cite{FISTA}} and PIDAL (Poisson image deconvolution by augmented Lagrangian) algorithm~\cite{Bioucas2010}, respectively. We applied both FISTA and PIDAL to perform reconstruction in Case~1 and compared them in Fig.~\ref{fig:fistavspista}. We assessed the image quality quantitatively using the mean-square-error (MSE) and structural similarity (SSIM)~\cite{wang2004image}. MSE is the mean-squared-difference between the reconstructed image and the ground truth, and SSIM measures the similarity between the reconstructed image and the ground truth. {Lower} MSE and {higher} SSIM mean better reconstruction. As shown in Table~\ref{table:fistavspista}, the image reconstructed with FISTA resulted in a lower MSE and higher SSIM, compared to PIDAL{, and PIDAL took approximately 20\% longer to run than FISTA.} Therefore, we used FISTA as the main image reconstruction algorithm.
\begin{figure}[!htbp]
\captionsetup{font=footnotesize}
    \centering
    \includegraphics[width=\linewidth]{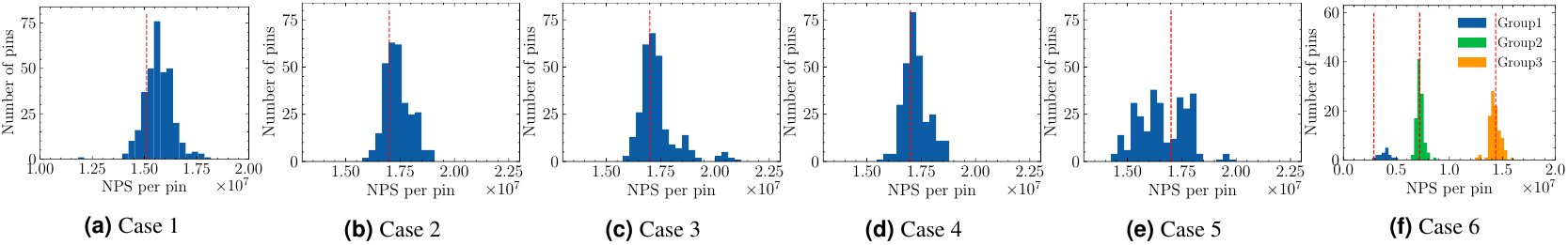}
    \caption{Activity distribution based on the inverse approach.{ The pin activity was estimated from each image by summing up the pixel values inside the present pins.} The ground truth is shown as the red line.}
    \label{fig: inv_act}
\end{figure}
\begin{table}[!htbp]
\captionsetup{font=footnotesize}
\centering
\caption{Comparison of MSE, SSIM, and computation time using FISTA and PIDAL.{ The computation time was measured on an Intel Core i9-7920X CPU.}}\label{table:fistavspista}
\resizebox{.35\linewidth}{!}{
    \begin{tabular}{|c|c|c|c|}
    \hline
    Method  & {MSE} & {SSIM} & Computation time (s) \\ \hline
    PIDAL & 42.05       & 0.52      & 231.68  \\ \hline
    FISTA & 41.45       & 0.59      & 183.57  \\ \hline
    \end{tabular}}
\end{table}

\begin{figure}[!htbp]
\captionsetup{font=footnotesize}
    \centering
    \includegraphics[width=\linewidth]{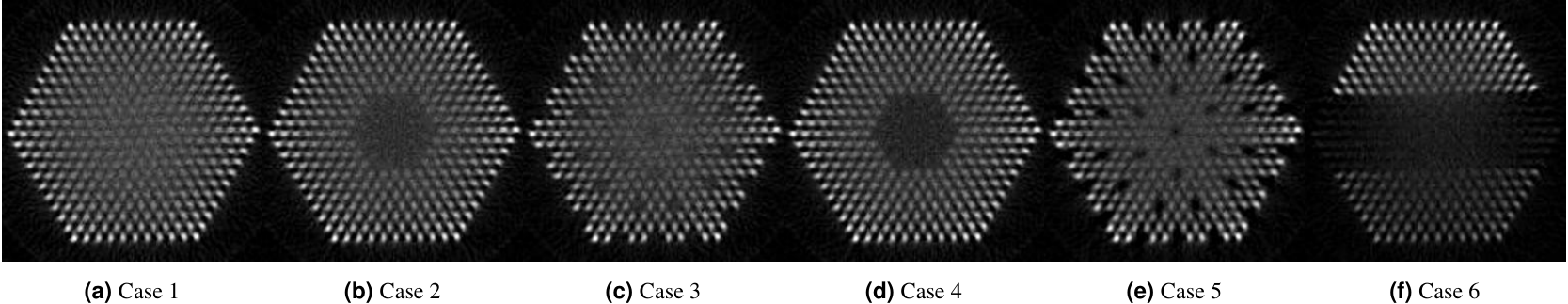}
    \caption{Image reconstructed using FBP.{ Severe image blurring occurred at the center of Case1, Case 3 and Case 5.}}
    \label{fig: fbp_recon}
\end{figure}
\begin{figure}[!htbp]
\captionsetup{font=footnotesize}
    \centering
    \includegraphics[width=\linewidth]{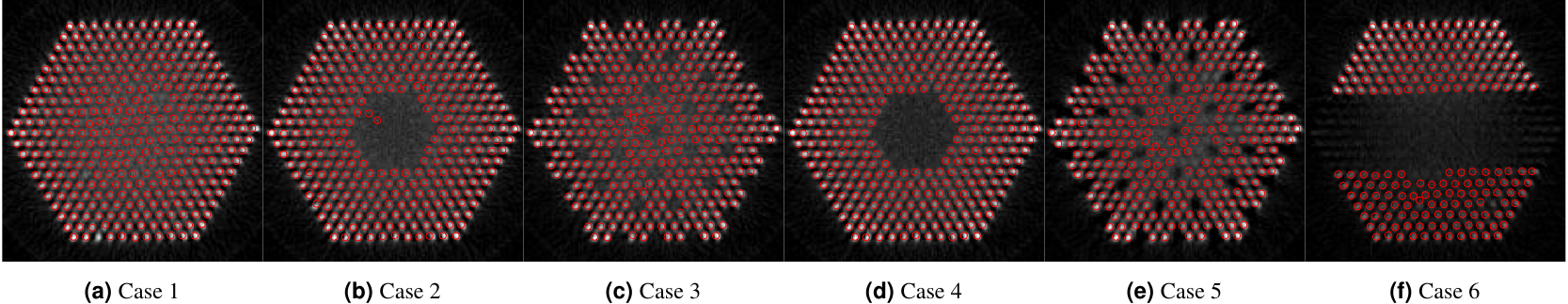}
    \caption{Pin identifications based on FBP. Identified pins are shown as red circles.{ Mis-classification of fuel pins appeared mostly at the center of Case 1, Case 3, and Case 5, where the hexagonal symmetry of the pin distribution was broken.}}
    \label{fig: fbp_pid}
\end{figure}
\begin{figure}[!htbp]
\captionsetup{font=footnotesize}
    \centering
    \includegraphics[width=\linewidth]{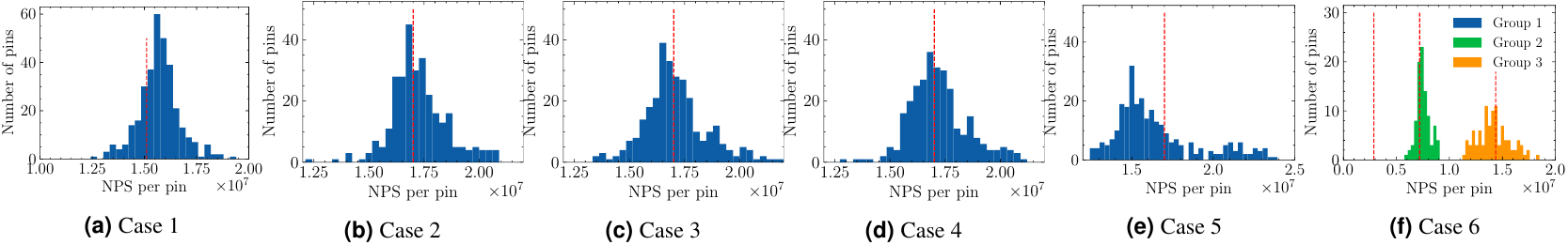}
    \caption{Activity distribution based on FBP.{ The pin activity was estimated by multiplying a normalization constant to the sum of all pixels inside the pin}.The ground truth is shown as the red line.}
    \label{fig: fbp_act}
\end{figure}
For comparison, we have also implemented the traditional FBP method to reconstruct the image, which is detailed in ``Methods-Image Reconstruction Methods" section. We then input these images to the neural network for pin identification, and estimated the pin activity levels. We compared the performance of the inverse (FISTA) and FBP approaches in terms of image quality, accuracy of pin identification and activity quantification. Fig.~\ref{fig: fbp_recon} shows the images reconstructed using FBP. Compared to Fig.~\ref{fig: fbp_recon}, images in Fig.~\ref{fig: inv_recon} demonstrated higher contrast and fewer reconstruction artifacts. The blurring at the center was significantly removed using the inverse approach. We calculated the MSE and SSIM for both sets of images, shown in Table~\ref{table:msessim}. We achieved significantly lower MSE and higher SSIM by using the inverse approach, which resulted in lower mis-classification rates of pin identification overall.{ The pin identification results based on FBP reconstruction are shown in Fig.~\ref{fig: fbp_pid}. In Fig.~\ref{fig: fbp_pid}a, Fig.~\ref{fig: fbp_pid}c, and Fig.~\ref{fig: fbp_pid}e, the blurring led to inaccurate pin localization around the center of FBP-reconstructed images, and the hexagonal structure was destroyed. In contrast, the hexagonal structure was accurately reproduced using the inverse problem approach, as shown in Fig.~\ref{fig: inv_pid}a-~\ref{fig: inv_pid}e}.{ Hence, given the same sinogram data, the inverse approach is expected to better distinguish \replaced{pins missing from the fuel assembly}{missed pins from the fuel assembly} and result in lower false alarm rates, compared to FBP. In Case~6, \added{we achieved the best MSE and SSIM, due to the smallest average pixel intensity in all cases. }\replaced{B}{b}oth FBP and inverse approach led to relatively high mis-classification rates due to the low activity of the pins in the middle of the assembly. However, we do not expect such a large variation of pin activities in an actual fuel assembly~\cite{jardine2005radiochemical}}.
\begin{table}[!htbp]
\captionsetup{font=footnotesize}
\centering
\caption{Comparison of image quality and mis-classification rates for the six simulated cases reconstructed using traditional FBP and the inverse (INV) approach. FP: false positive{ (missing pins mis-classified as present)} rate; FN: false negative{ (present pins mis-classified as missing)} rate.}\label{table:msessim}
\resizebox{.5\linewidth}{!}{
    \begin{tabular}{|c|c|c|c|c|c|c|c|c|}
    \hline
    \multirow{3}{*}{Case} & \multicolumn{2}{c|}{MSE}                    & \multicolumn{2}{c|}{SSIM}                   & \multicolumn{4}{c|}{Mis-classification rate}        \\ \cline{2-9} 
                          & \multirow{2}{*}{FBP} & \multirow{2}{*}{INV} & \multirow{2}{*}{FBP} & \multirow{2}{*}{INV} & \multicolumn{2}{c|}{FBP} & \multicolumn{2}{c|}{INV} \\ \cline{6-9} 
                          &                      &                      &                      &                      & FP          & FN         & FP         & FN          \\ \hline
    1                     & 96.62                & 41.45                & 0.16                 & 0.59                 & 0      & 1.51\%     & 0     & 0           \\ \hline
    2                     & 97.01                & 41.76                & 0.16                 & 0.59                 & 0.68\%      & 0     & 0     & 0           \\ \hline
    3                     & 95.32                & 40.69                & 0.16                 & 0.56                 & 0.34\%      & 0     & 0          & 0           \\ \hline
    4                     & 96.63                & 38.43                & 0.16                 & 0.64                 & 0           & 0     & 0          & 0           \\ \hline
    5                     & 88.70                & 37.38                & 0.17                 & 0.60                 & 0      & 3.40\%    & 0          & 0      \\ \hline
    6                     & 83.35                & 36.33                & 0.21                 & 0.63                 & 0      & 35.03\%    & 0     & 27.55\%     \\ \hline
    \end{tabular}}
\end{table}

We estimated the activity of each pin by summing the pixels on the reconstructed images and created a histogram of pin activities for each case, shown in Fig.~\ref{fig: inv_act} and Fig.~\ref{fig: fbp_act}. {It should be noted that for the FBP reconstruction, there is no definite relationship between the pixel sum and the absolute activity. Appropriate normalization constant needs to be applied to convert the pixel sum into activity, which is not always possible in an actual inspection. In this case, {we calculated the ratio between the mean of the pixel sum distribution and the true activity for Cases 2-5. We then used the average of these ratios as the normalization constant and applied it to all cases to convert the pixel sum to absolute activity.} In contrast, normalization is not needed for the inverse reconstruction because the response matrix has already been normalized per unit source particle. }Compared to Fig.~\ref{fig: fbp_act}, the pin activity distributions in Fig.~\ref{fig: inv_act} were narrower. We calculated the relative error compared to the ground truth and the standard deviation of pin activity, as shown in Table~\ref{table:act_esti}. We obtained smaller standard deviations by using the inverse approach in all cases. The relative error was negligible compared to the ground truth, except for the low activity pins in Group 1 of Case 6.
\begin{table}[!htbp]
\captionsetup{font=footnotesize}
\centering
\caption{Comparison of activity estimations.{ The mean NPS per pin is the mean of the activity distribution; relative error is the relative difference between the mean activity and ground truth; standard deviation refers to the standard deviation of the pin activity distribution. For FBP in Group 1 of Case 6, no data is shown because no present pin is identified.}}\label{table:act_esti}
\resizebox{.6\linewidth}{!}{
    \begin{tabular}{|c|c|c|c|c|c|c|c|c|}
    \hline
    \multicolumn{2}{|c|}{\multirow{2}{*}{Case}} & \multicolumn{3}{c|}{{Mean} NPS per pin ($\times 10^7$)} & \multicolumn{2}{c|}{Relative Error (\%)} & \multicolumn{2}{c|}{Standard deviation (\%)} \\ \cline{3-9} 
    \multicolumn{2}{|c|}{}                      & Truth      & FBP         & INV         & FBP                 & INV                & FBP                   & INV                  \\ \hline
    \multicolumn{2}{|c|}{1}                     & 1.51       & 1.57        & 1.57       & 3.91             & 3.74            & 6.16                & 4.36              \\ \hline
    \multicolumn{2}{|c|}{2}                     & 1.70       & 1.72        & 1.74       & 1.39              & 2.21            & 7.57               & 3.44               \\ \hline
    \multicolumn{2}{|c|}{3}                     & 1.70       & 1.71       & 1.73       & -0.68            & 2.02             & 8.54                & 5.09               \\ \hline
    \multicolumn{2}{|c|}{4}                     & 1.70       & 1.72       & 1.73       & 1.12             & 1.46            & 7.71               & 3.33              \\ \hline
    \multicolumn{2}{|c|}{5}                     & 1.70       & 1.65       & 1.66       & -2.91            & -2.25           & 15.75               & 6.85               \\ \hline
    \multirow{3}{*}{6}         & Group 1        & 0.29       & {N.A.}     & 0.39        & {N.A.}          & 35.73           &{N.A.}                    & 14.22                \\ \cline{2-9} 
                               & Group 2        & 0.72       & 0.75        & 0.72        & 4.69                & 0.54               & 8.25                  & 4.64                 \\ \cline{2-9} 
                               & Group 3        & 1.44       & 1.42        & 1.43        & -1.25               & -0.59            & 10.68             & 3.70                 \\ \hline
    \end{tabular}}
\end{table}

{Compared to $\mathrm{UO}_2$ fuel rod, the cobalt rod simulated in this work resulted in less attenuation because of its lower density and atomic number. Nevertheless, this work describes a general approach for correcting the gamma-ray attenuation and down-scattering in the energy window of interest and it can be easily adapted to $\mathrm{UO}_2$ fuel assemblies by updating the atomic composition and gamma-ray source term in the simulation.}
{In this work, we assumed that the detector \replaced{performance}{gain} is \replaced{the same}{uniform} in the simulation of the sinogram and response matrix, which is not true for the real PGET device. The correction for \replaced{varying detector performances}{non-uniform detector gains} across the detectors, either at the hardware or software level, is crucial to obtain an accurate estimation of pin activity. When identifying drifting detectors in post-processing, e.g., whose response is anomalously different compared to neighbour detectors, a simple approach consists in neglecting their response. In a preliminary analysis on simulated data, we replaced the response of an increasing number of detectors in the sinogram with null arrays. We found that if the number of ``neglected'' detectors is below ten in the PGET setup, the reconstructed image is negligibly affected by not including their response. Future work will be needed to study the robustness of the software suite as a function of a systematic bias in the detector \replaced{performance}{gain}.}

\section*{Conclusion}

{In this work, we have implemented a full set of software, which is able to reconstruct cross-sectional images of mockup fuel assemblies acquired by a simulated PGET system, identify missing fuel pins, and estimate fuel pin activities based on the reconstructed image. We have developed a linear forward model that accounts for the scattering of gamma rays in the assembly to accurately characterize the response matrix of the PGET system. The image reconstruction was formulated into a linear inverse problem by modeling the observation noise as Gaussian, which was solved using FISTA. 
The reconstructed image was fed to a convolutional neural network to automatically identify the present pins and determine their centroids. Compared to the FBP approach, the inverse problem approach resulted in over 50\% lower MSE and 200\% higher SSIM, and consequently lower mis-classification rates in pin identification in all cases. Based on the pin identification results, we estimated the pin activity by summing up the pixel values around the centroid inside the pin radius on the image. Compared to the FBP, the inverse approach resulted in smaller standard deviations of pin activity in all cases, with negligible bias with respect to the ground truth. The proposed inverse approach\added{ to reconstruct a fuel pin cross section, identify fuel pins and calculate their activity }took approximately 8 minutes to run on an Intel Core i9-7920X CPU without parallelization. We are currently improving the algorithm to allow automatic classification of different activity groups in a real fuel assembly.}

\section*{Methods}
\subsection*{Monte Carlo Simulation of the PGET based on MCNP }
\begin{figure}[!htbp]
\captionsetup{font=footnotesize}
 \centering
  \includegraphics[width=.5\linewidth]{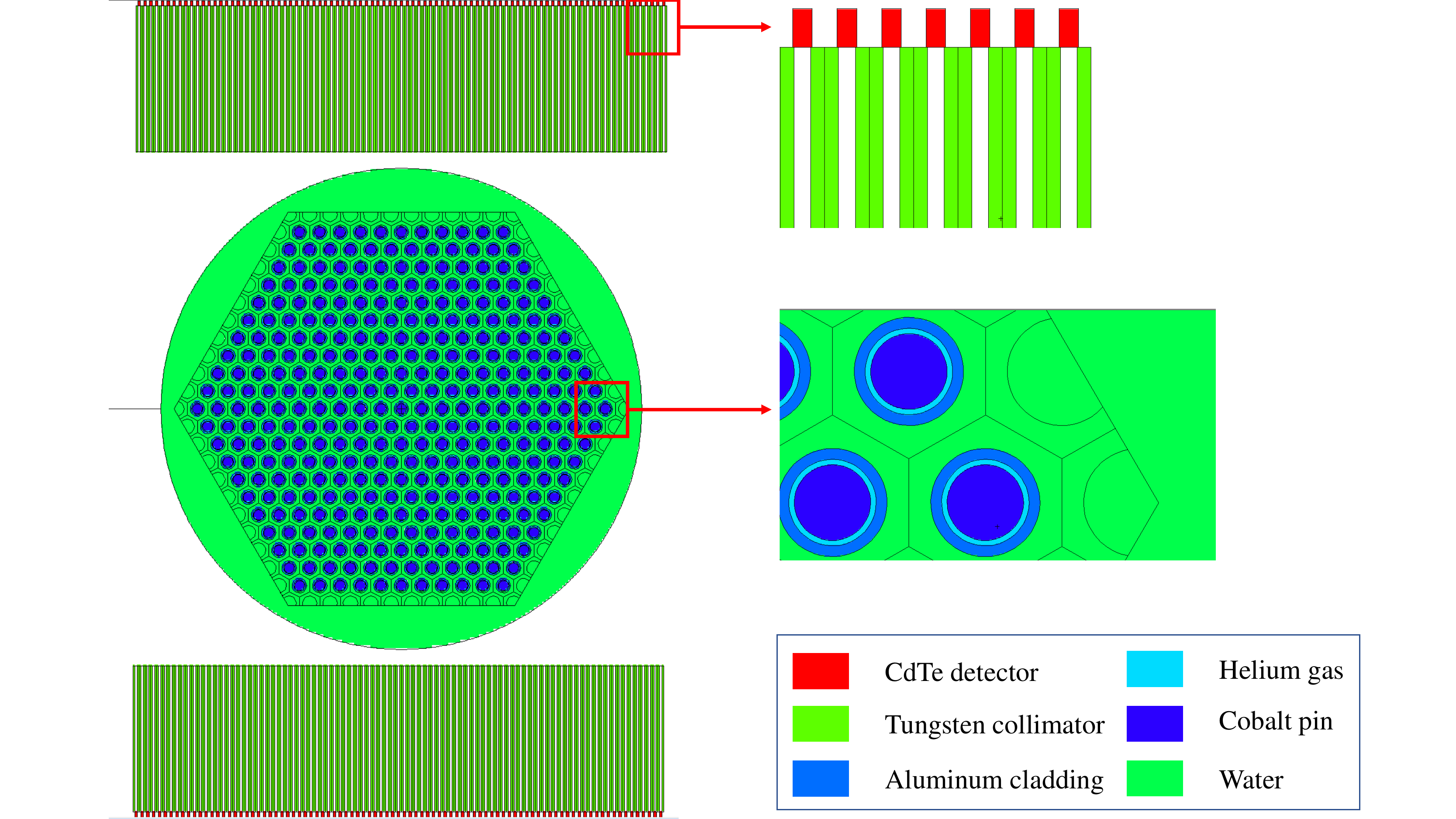}
  \caption{The cross-sectional view of the MCNP model in Case 1.{ The hexagonal fuel assembly contains 331 ${}^{60}$Co rods and is submerged in a water cylinder of 33 cm diameter. The fuel pin is simulated as a ${}^{60}$Co cylinder of 7~mm diameter, coated with 1~mm thick aluminum cladding. Gamma rays emitted by the ${}^{60}$Co rods are detected by two collimated CdTe detector arrays located on two sides of the fuel assembly, each encompassing 91 detectors. The tungsten collimator between the detector array and fuel assembly is 10~cm long and has an aperture of 1.5~mm. Each CdTe detector is closely attached to the collimator and is of size $0.175\times0.35\times0.35{\text{ cm}}^{3}$. The distance between two neighbouring detectors is 4~mm. The bottom array is shifted to left by 2~mm, with respect to the top one~\cite{honkamaa2014prototype,white2018application}. The detector arrays rotate clockwise and scan the fuel bundle in steps of 1$\degree$, which generates a $182\times360$ sinogram.}}
  \label{fig:PGET}
\end{figure}

Fig.~\ref{fig:PGET} shows the cross-sectional view of the MCNP model of Case 1. {In Case 4 and 5, depleted uranium pins were used, where Co was replaced by  {0.20\% enrichment }UO${}_{2}$. }{We used the F4 tally as the detector response model in the MCNP simulation and the response matrix calculation. The F4 tally estimates the average photon flux in the detector cell by summing the track lengths of all particles in the 700-1500 keV energy window~\cite{mcnp2003general}. The absolute activity measurement can be obtained using the proposed method as long as the simulated quantity in the response matrix is consistent with the simulated response to an unknown inspected fuel bundle. When applying the proposed technique to experimental data, one would want to validate the simulated detector response with the measured one, to properly incorporate specific detector's properties, such as its energy resolution and time response, in the simulated model.}

\subsection*{Image Reconstruction Methods}
In this section, we describe in detail the two image reconstruction methods implemented in this work: the FBP and linear inverse problem method. We applied both methods to reconstruct images from the sinograms and compared their performance.
\subsubsection*{Filtered Back-Projection}\label{sec:fbp}
\begin{figure}[!htb]
\captionsetup{font=footnotesize}
 \centering
    \includegraphics[width=.4\linewidth]{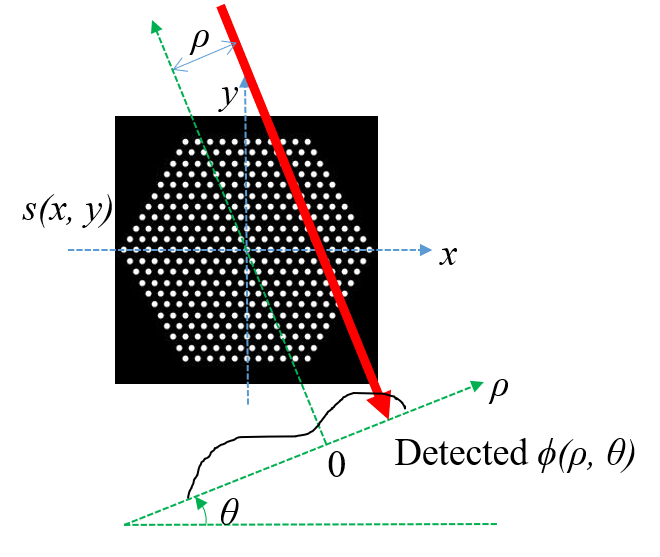}
    \caption{{The tomography measurement of the fuel assembly at an angle $\theta$. The counts of detector $(\rho,\theta)$ can be approximated by the integral of the source distribution $s(x, y)$ along the red line that passes through the detector center.}}
    \label{fig:fbp_geometry}
\end{figure}

Fig.~\ref{fig:fbp_geometry} shows the tomography measurement of the fuel assembly at angle $\theta$. The FBP method relies on two approximations: first, we neglect the attenuation and scattering of gamma rays in the system; second, for pins at different locations, we assume that the geometric efficiencies are the same. Under these approximations, the sinogram $\boldsymbol{\phi}(\rho,\theta)$ can be simplified as the integral of the source distribution $\mathbf{s}(x,y)$ along the red line passing through the detector center, \textit{i.e.},  
\begin{equation}
\begin{aligned}
    \boldsymbol{\phi}(\rho,\theta) = \iint_{\mathbb{R}^2} \mathbf{s}(x,y) \delta(x \cos\theta + y \sin\theta-\rho) dxdy.
\end{aligned}\label{eq:fbp}
\end{equation}
Eq.~\eqref{eq:fbp} is the standard forward model in X-ray tomography imaging. The classical inverse operation from the sinogram to the source is the filtered back-projection, which is also a standard imaging algorithm in X-ray imaging~\cite{prince2006medical} and will not be detailed here.

\subsubsection*{Inverse Problem Approach}

The inverse reconstruction approach is based on the conversion of the image reconstruction problem into a linear inverse problem. Let the vectorized image of the fuel assembly be $\mathbf{s}$, which contains $N$ unknown pixel values, and the vectorized simulated sinogram of size $M=65,520$ ($182 \times 360$) be $\boldsymbol{\phi}$. The inverse approach relies on the assumption that $\boldsymbol{\phi}$ is a linear function of $\mathbf{s}$:
\begin{equation}
\begin{aligned}
    \boldsymbol{\phi} = \mathbf{A}\mathbf{s} + \mathbf{n},
\end{aligned}\label{eq:inverse1}
\end{equation}
where $\mathbf{A}$ is the system response matrix of size $M \times N$, and $\mathbf{n}$ models random observation noise, assumed to be isotropic Gaussian distributed. Physically, the $i$-th column of the response matrix $\mathbf{A}$ is the vectorized sinogram corresponding to a pin distribution with unit activity in pixel $i$ but zero elsewhere. Accurate determination of the system response matrix is crucial to obtain a high-quality image and avoid systematic bias in pin identification and activity estimation. 
The analytical derivation of the response matrix is discussed in the next section.

Given the simulated sinogram and system response matrix $\mathbf{A}$, we can reconstruct the image by solving the following equation:
\begin{equation}\label{eq:FISTA}
        \hat{\mathbf{s}} =  \operatorname*{arg\,min}_{{s_i\geq0, \forall i}} \quad \left[\frac{1}{2}\|\boldsymbol{\phi} -\mathbf{A}\mathbf{s}\|_2^2  + \lambda \|\mathbf{s}\|_1 \right] = \operatorname*{arg\,min}_{{s_i\geq0, \forall i}} \quad \left[\frac{1}{2}(\boldsymbol{\phi} -\mathbf{A}\mathbf{s})^T(\boldsymbol{\phi} -\mathbf{A}\mathbf{s})  + \lambda \sum_{i=1}^N|\mathbf{s}_i| \right]
\end{equation}
where the first term is the data-fidelity term assuming Gaussian noise, and the second term is a regularization term acknowledging that the fuel pin is sparsely distributed. The regularization parameter $\lambda$ is chosen based on the noise level. To solve Eq.~\eqref{eq:FISTA}, we have implemented the fast iterative shrinkage-thresholding algorithm (FISTA)~\cite{FISTA}.

As an alternative, we can model the observation noise by Poisson noise and the data-fidelity term is changed accordingly~\cite{Bioucas2010,lefkimmiatis2013poisson} as follows
\begin{equation}\label{eq:PISTA}
        \hat{\mathbf{s}} =  \operatorname*{arg\,min}_{{s_i\geq0, \forall i}} \quad \left[\sum_{j=1}^{M}([\mathbf{A}\mathbf{s}]_j - \boldsymbol{\phi}_j \log[\mathbf{A}\mathbf{s}]_j) + \lambda \|\mathbf{s}\|_1 \right].
\end{equation}
To solve Eq.~\eqref{eq:PISTA}, we used the PIDAL algorithmic structure described in \cite{Bioucas2010}. 

\subsubsection*{Computation of the Response Matrix}
The calculation of the system response matrix $\mathbf{A}$ involves the calculation of detector response to each pixel, with each response being one column in the matrix. For inner pixels, the contribution from scattered photons is non-negligible due to the high attenuation. To account for this, we have implemented an algorithm that combines stochastic photon transport with deterministic collimator-detector modeling to accurately calculate the system response matrix.

{For real fuel assemblies, no prior information of pin distribution and pin radius will be given. It is therefore necessary to first estimate these quantities before running the Monte Carlo simulation. {As discussed in the “Results-Inverse approach” section, }this is done by performing a rough image reconstruction using our previously developed model, where{ uniform attenuation map is used and}{ no scattering effect is considered}~\cite{fanginmm2020}. Based on the reconstructed image, one could determine the pin locations and pin radius, and create a material map,  where the material is assumed to be cobalt inside the pin, and water outside.}

Next, we simulate the travel of photons inside the fuel assembly using Monte Carlo.{ Two sets of grids are generated. The fine grid has a pixel size of $0.5\times0.5$~mm${}^{2}$, while the coarse grid has a pixel size of $2\times2$~mm${}^{2}$. {We first discretize the region of interest on the fine grid and calculate the response to the pixels whose center points are inside the pin.} Then we calculate the response to a pixel on the coarse grid by summing the responses to the corresponding $4\times 4$ pixels on the fine grid. \replaced{Based on the pin identification result in the previous step, the potential pin-present region are pixelated into 5,329 coarse pixels. We iterate over all these pixels}{. We iterate over all coarse pixels} and form the response matrix. In this way, we are able to reduce the pixelization error introduced during the discretization of the pins, without increasing the final response matrix size.}

The response to each {fine }pixel {whose center point} is inside the pin is calculated in the following way. A photon $(\vec{r_0}, \vec{\Omega_0}, E_0, W_0)$ is created, with the initial position $\vec{r_0}$ uniformly sampled inside the pixel, the initial moving direction $\vec{\Omega_0}$ uniformly sampled in $4\pi$ space, the initial energy $E_0$ sampled from the source energy spectrum, and the initial weight \added{$W_0$ }assigned to 1. We track the photon inside the PGET using the delta-tracking algorithm~\cite{woodcock1965techniques} and determine the position where the photon interacts with medium. We force the interaction to be Compton scattering by multiplying its weight with the probability that the interaction is Compton scattering,

\begin{equation}
    W_{i} = W_{i-1} \frac{\mu_{\text{sc}}(\vec{r}_{i-1}, E_{i-1})}{\mu(\vec{r}_{i-1},E_{i-1})}
\end{equation}
where $\mu_{\text{sc}}(\vec{r}_{i-1}, E_{i-1})$ and ${\mu(\vec{r}_{i-1}, E_{i-1})}$ are the Compton scattering attenuation coefficient and total attenuation coefficient at the $i$-th interaction site, respectively. A scattering angle is sample based on Kahn's rejection algorithm~\cite{kahn1954applications}, and the photon energy $E_{i}$ and moving direction $\vec{\Omega_i}$ are updated accordingly. This process is repeated until the photon escapes the fuel assembly or the photon energy is below the detection threshold. A copy of the photon is saved at its creation site and each interaction site to a sub-projection map. 

\begin{figure}[!htbp]
\captionsetup{font=footnotesize}
 \centering
  \includegraphics[width=.5\linewidth]{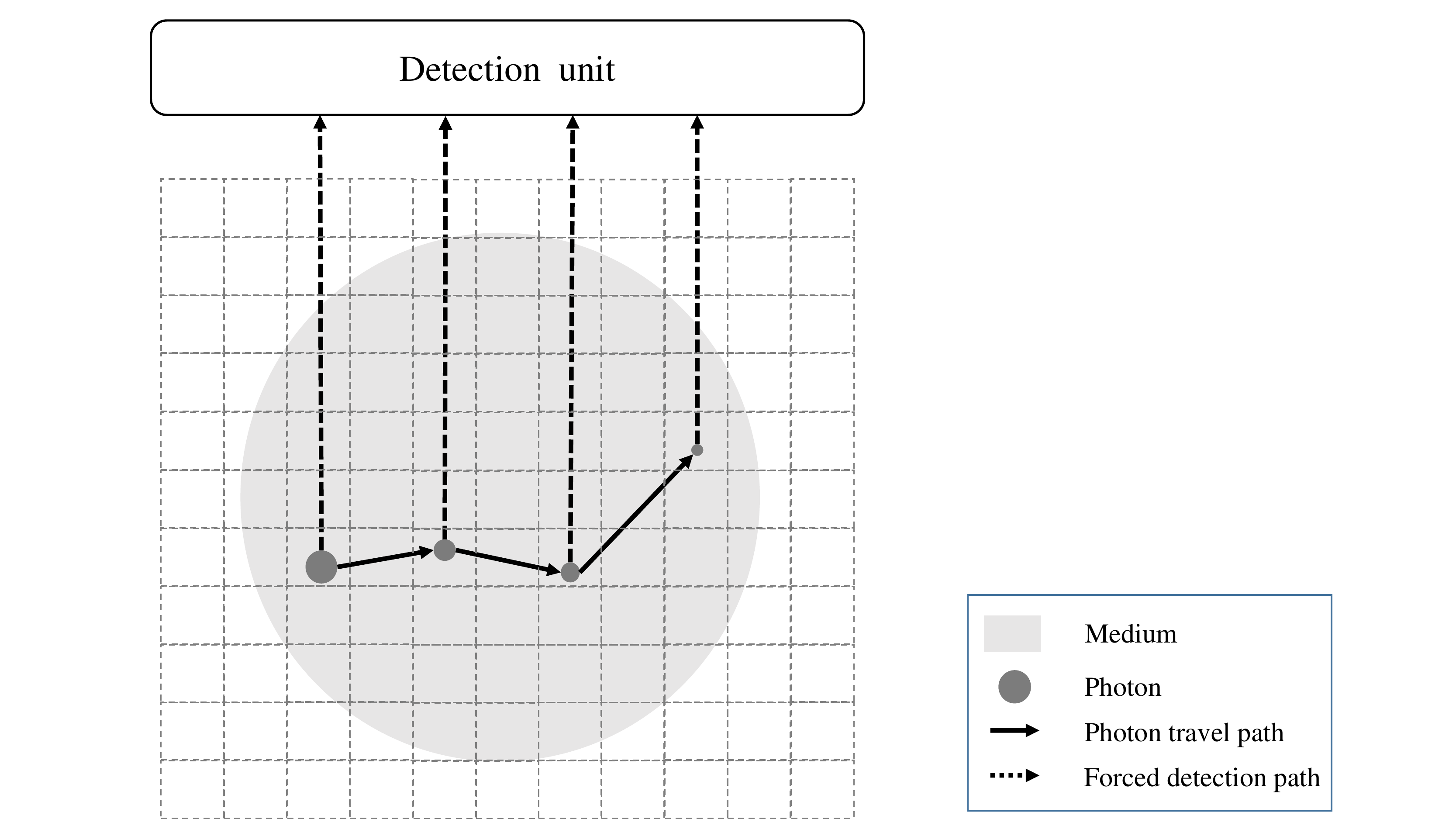}
  \caption{Illustration of the CFD{ technique}.{ A photon shown as the gray disk is created on the left and travels in the medium along the solid arrow. The disk radius represents the photon energy. Upon it creation and each interaction with the medium, the photon is forced to be scattered along the dash arrow and to be detected by the detector with a certain probability. }}
  \label{fig:cfd}
\end{figure}

 {Once the sub-projection map is obtained, we apply the convolutional forced detection (CFD) technique to calculate the detector response. }CFD is a widely-used variance reduction technique for X-ray downscatter simulation in single photon emission computed tomography (SPECT){, which has been shown to be 50-100 times faster than conventional forced detection technique~\cite{hutton2011review} and several thousand times faster than full-3D Monte Carlo simulation~\cite{beekman2001efficient}}. Fig.~\ref{fig:cfd} illustrates its principle. In CFD, the photon is forced to travel perpendicularly to the detector plane upon its creation or interaction with medium to effectively simulate all the possible spatial distributions of particle interactions and consequent detection events. For an unscattered photon, the contribution to the F4 tally in the detector cell is given by 
\begin{equation}\label{eq:unsccfd}
    W\times \frac{\Delta\Omega}{4\pi}\times e^{-\int\mu(\vec{r},E) ds} \times \frac{T_l}{V}
\end{equation}
and for a scattered photon, the contribution is 
\begin{equation}\label{eq:sccfd}
    W \times \frac{\sigma (\theta) \Delta\Omega}{2\pi \int_{0}^{\pi}\sigma (\theta')\sin(\theta')d\theta' } \times e^{-\int\mu(\vec{r},E(\theta)) ds} \times \frac{T_l}{V}
\end{equation}
In Eq.~\eqref{eq:unsccfd} and \eqref{eq:sccfd}, $W$ is the photon weight, $\Delta\Omega$ is the solid angle subtended by the detector\added{ in steradian}, $\sigma(\theta)$ is the differential cross-section for Compton scattering, $\mu(\vec{r}, E)$ is the total attenuation coefficient along the CFD path, and the last term stands for the contribution to F4 tally, \textit{i.e.}, average track length in detector cell normalized by detector volume.

Due to the non-ideal collimation, a photon will contribute counts not only to the detector in the same column, but also to the neighbouring ones. This phenomenon is characterized by the depth-dependent point spread function (PSF) of the collimator,  \textit{i.e.}, the photon count distribution on the detection plane when a point source is imaged. PSF depends on the vertical distance $z$ between the photon and the detector array~\cite{de2001acceleration}:
\begin{equation}
\begin{gathered}
    {PSF_z(x) = \replaced{\frac{2\sqrt{\ln(2)}}{\sqrt{\pi}\mathrm{FWHM}(z)}\exp(\frac{-4\ln{2}\times x^2}{\mathrm{FWHM}(z)^2})}{\frac{1}{\sqrt{2\pi}{\sigma}(z)}\exp(\frac{-x^2}{2\sigma(z)^2})}, \replaced{\mathrm{FWHM}}{\sigma}(z)  = b + \frac{b}{l_{\text{eff}}}z, {l_{\text{eff}}} = l - \frac{2}{\mu}}
\end{gathered}
\end{equation}
where $b$ is the collimator aperture, $l$ is the collimator length, $\mu$ is the total attenuation coefficient in the collimator\added{, and $\mathrm{FWHM}$ is the width of the collimator PSF}.

To calculate the response to a pin at an angle $\theta$, the photons saved in the sub-projection map are first rotated by an angle $-\theta$. We then sum the weights of the photons in the same pixel on the sub-projection map and calculate their contribution to the detector response based on Eq.~\eqref{eq:unsccfd} or \eqref{eq:sccfd}, which creates a projection map. We then convolve each row in the projection map with the PSF function and sum all rows up to get the response at this angle. Next, we increment the angle $\theta$ and calculate the response at the next angle. In this way, photon tracking in the fuel assembly is done only once, which saves computation time.

We simulated the travel of 6,400 photons to create the sub-projection map for one pixel. It took approximately 4 minutes to calculate the response matrix containing 5,329 pixels using CFD on an Intel Core i9-7920X CPU.


\subsection*{Pin Localization}
Images reconstructed using both approaches are input to a convolution neural network, referred to as U-net~\cite{ronneberger2015u}, to perform pin identification. {The experimental data provided by IAEA within the framework of a technology open challenge~\cite{iaea2019challenge} were used for the training of the U-net. The training data consisted of the FBP images reconstructed from the measured sinograms of mock-up assemblies of 12 different geometries with a variable number of ${}^{60}$Co pins, and examples of the reconstructed images of these assemblies can be found in~\cite{iaea2019challenge}.}{ The U-net was trained to minimize a binary crossentropy loss between the predicted pin masks and ground truth. For each image, we provided four randomly rotated ones as input to the network, for data augmentation. Data augmentation allowed us to generate new training instances from existing ones, artificially boosting the size of the training set. The network was trained for 10,000 epochs, \textit{i.e.}, complete iteration over the whole dataset, using a learning rate of $10^{-5}$ and an Adam optimizer~\cite{kingma2014adam}. We found that $10^{-5}$ is the learning rate for which the slope of the crossentropy loss is maximized, allowing to reach the minimum. The training was early stopped when the loss reached its minimum value.
} The trained U-net can extract visible structures from the input image and output a mask based on the extracted features. Bright regions on the mask with areas above a certain threshold are identified as present pins, and their centroids are calculated as the coordinates of the center of each pin.


\section*{Acknowledgements}
This work was funded in part by the Nuclear Regulatory Commission Faculty Development Grant number 31310019M0011\added{ and the Consortium for Verification Technology under Department of Energy National Nuclear Security Administration award number DE-NA0002534}. This work was also supported by the Royal Academy of Engineering under the Research Fellowship scheme RF201617/16/31.  

\section*{Author contributions statement}

 {M.F. developed the forward model and analyzed the data, Y.A. implemented the FISTA and PIDAL algorithms, D.D.L. and M.S. developed and trained the neural network, A.D.F. implemented the FBP and the MC model, conceived and oversaw the project. M.F., Y.A., A.D.F., and M.S. wrote the manuscript. All authors reviewed the manuscript. }

\section*{Additional information}

The authors declare no competing interests. 
Software not already available online will be made publicly and freely available on the authors' website once this manuscript is accepted for publication. 

\end{document}